%% file: main.tex
\documentclass[acmsmall]{acmart} 

\renewcommand\footnotetextcopyrightpermission[1]{} 
\AtBeginDocument{%
  \providecommand\BibTeX{{%
    \normalfont B\kern-0.5em{\scshape i\kern-0.25em b}\kern-0.8em\TeX}}}

\usepackage{tabu}
\usepackage{colortbl,hhline}
\usepackage{multirow}
\usepackage{graphicx}

\setcopyright{rightsretained}
\acmJournal{PACMHCI}
\acmYear{2024} \acmVolume{8} \acmNumber{CSCW2}
\acmArticle{434} \acmMonth{11}
\acmDOI{10.1145/3686973}

\begin{document}

\title{Articulation Work and Tinkering for Fairness in Machine Learning}

\author{Miriam Fahimi}
\email{miriam.fahimi@aau.at}
\orcid{0000-0002-0619-3160}
\affiliation{%
  \institution{Digital Age Research Center, University of Klagenfurt}
  \city{Klagenfurt}
  \country{Austria}
}

\author{Mayra Russo}
\authornote{The second and third authors contributed equally to this research.}
\email{mrusso@l3s.de}
\orcid{0000-0001-7080-6331}
\affiliation{%
  \institution{L3S Research Center, Leibniz University Hannover}
  \city{Hannover}
  \country{Germany}
}

\author{Kristen M. Scott}
\authornotemark[1]
\email{kristen.scott@kuleuven.be}
\orcid{0000-0002-3920-5017}
\affiliation{%
  \institution{KU Leuven, Leuven.AI}
  \city{Leuven}
  \country{Belgium}
}

\author{Maria-Esther Vidal}
\email{maria.vidal@tib.eu}
\orcid{0000-0003-1160-8727}
\affiliation{%
  \institution{TIB Leibniz Information Center for Science and Technology, L3S Research Center \& Leibniz University of Hannover}
  \city{Hannover}
  \country{Germany}
}

\author{Bettina Berendt}
\email{berendt@tu-berlin.de}
\orcid{0000-0002-8003-3413}
\affiliation{%
  \institution{TU Berlin, Weizenbaum Institute, and KU Leuven}
  \city{Berlin and Leuven}
  \country{Germany and Belgium}
}

\author{Katharina Kinder-Kurlanda}
\email{katharina.kinder-kurlanda@aau.at}
\orcid{0000-0002-7749-645X}
\affiliation{%
  \institution{Digital Age Research Center, University of Klagenfurt}
  \city{Klagenfurt}
  \country{Austria}
}

\renewcommand{\shortauthors}{Fahimi, Russo, Scott, Vidal, Berendt and Kinder-Kurlanda}
\renewcommand{\shorttitle}{Articulation Work and Tinkering for Fairness in ML}
\begin{abstract}
The field of fair AI aims to counter biased algorithms through computational modelling. However, it faces increasing criticism for perpetuating the use of overly technical and reductionist methods. As a result, novel approaches appear in the field to address more socially-oriented and interdisciplinary (SOI) perspectives on fair AI. In this paper, we take this dynamic as the starting point to study the tension between computer science (CS) and SOI research. By drawing on STS and CSCW theory, we position fair AI research as a matter of `organizational alignment’: what makes research `doable’ is the successful alignment of three levels of work organization (the social world, the laboratory, and the experiment). Based on qualitative interviews with CS researchers, we analyze the tasks, resources, and actors required for doable research in the case of fair AI. We find that CS researchers engage with SOI research to some extent, but organizational conditions, articulation work, and ambiguities of the social world constrain the doability of SOI research for them. Based on our findings, we identify and discuss problems for aligning CS and SOI as fair AI continues to evolve. 
\end{abstract}




\maketitle

\input{sections/01introduction.tex}

\input{sections/02background.tex}
\input{sections/03case.tex}
\input{sections/04method.tex}

\input{sections/05findings.tex}
\input{sections/06discussion.tex}
\input{sections/07limitations}
\input{sections/08conclusion}

\begin{acks}
This work has received funding from the European Union’s Horizon 2020 research and innovation programme under Marie Sklodowska-Curie Actions (grant agreement number 860630) for the project ``NoBIAS - Artificial Intelligence without Bias'' and from the Flemish Government (AI Research Program). Maria-Esther Vidal is partially supported by Leibniz Association, program ``Leibniz Best Minds: Programme for Women Professors'', project TrustKG-Transforming Data in Trustable Insights; Grant P99/2020. This work reflects only the authors' views, and the European Research Executive Agency (REA) is not responsible for any use that may be made of the information it contains.
\end{acks}

\bibliographystyle{ACM-Reference-Format}
\bibliography{CSCW}

\received{January 2024}
\received[revised]{April 2024}
\received[accepted]{May 2024}

\end{document}

%% file: sections/01introduction.tex
\section{Introduction}
\label{sec:intro}
Fair artificial intelligence (fair AI) has emerged as a novel research field for computer scientists and technologists to devise algorithmic interventions. The new field, in broad terms, is concerned with research that supports the development of trustworthy, responsible, ethical artificial intelligence (AI) \cite{10.1145/3306618.3314289, whittlestone_role_2019}, by proposing novel methods to incorporate fairness notions into these systems, among other types of interventions. 

As of the present moment, fair AI finds itself at an intermediary phase of disciplinary evolvement \cite{heilbron_regime_2004, merz_local_2016}. Notably, fair AI is navigating internal dissent and external challenges about its future orientation \cite{weinberg_rethinking_2022, miceli_studying_2022}. This dynamic is marked by the interplay of two contrasting research paradigms: one rooted in computer science (CS), the origin discipline of fair AI, and another one that is more socially-oriented and interdisciplinary (SOI) (e.g. \cite{birhane_values_2022, benbouzid_fairness_2023, selbst_fairness_2019}). On one side, there is a defense of mathematically rigorous fairness approaches by CS researchers\footnote{{\href{https://www.youtube.com/watch?v=g-z84_nRQhw}{Dwork 2018. The Emerging Theory of Algorithmic Fairness. Microsoft Research (\textit{accessed via YouTube, 2024}).}}}, and an urgent call to not discredit them\footnote{{\href{https://www.youtube.com/watch?v=gZrZwF3XDBw}{Lidgett 2021. In Praise of Flawed Mathematical Models. ACM FAccT Conference (\textit{accessed via YouTube, 2024}).}}}, particularly in light of the increasing presence of AI technologies. On the other side, extensive critiques persistently challenge CS-oriented fair AI research. This criticism often stems from the perception that these approaches view solutions to discrimination and social inequality through a techno-optimistic lens, relying on computational methods without sufficient engagement with the reality of a socially stratified and diverse society \cite{davis_algorithmic_2021}. Scholars from various disciplines \cite{davis_algorithmic_2021,wachter_why_2020,10.1145/3290605.3300760} have voiced these critiques, contributing to the formation of the SOI paradigm. Boundaries are not always clear-cut, including the case of the boundaries between CS and SOI paradigms, and thus it should be noted that some prominent CS researchers have played a key role in highlighting algorithmic discrimination \cite{buolamwini_gender_2018,oneil2016WeaponsMathDestruction} and offering self-critical perspectives  \cite{narayanan2022limits,ruggieri_can_2023}. Additionally, \citet{abebe_roles_2020} highlights how CS can serve as a constructive ally in pursuing social change (an SOI-oriented goal).

In this paper, we suggest that the persisting gap between CS research and SOI critiques originates not (only) from epistemic and disciplinary tension, but is also rooted in organizational alignment. With this, we mean the alignment of research across three levels of work organization. According to theories in science and technology studies (STS) and computer-supported cooperative work (CSCW), research becomes only `doable' when the three levels of work organization, e.g. the experiment, the laboratory, and the social world, are successfully aligned \cite{fujimura_constructing_1987, strauss_work_1985}. As a result, researchers must spend time and effort in articulation work to achieve this alignment across all organizational levels. Further, researchers must gather the necessary resources, technologies, equipment, and people, and align them, to do their research. Departing from this theoretical lens, and motivated by our own experiences of the tensions between the CS and SOI paradigms, this study sets out to understand CS researchers' perceptions and challenges of alignment with SOI in the case of fair AI.

To address our inquiry, we carried out an interview study with ten senior CS researchers based in different research centers and universities across Europe, who are actively working on fair AI research. Most of the interview participants were also collaborators of a European Union (EU)-funded project that focused on research and development of novel interdisciplinary methods for fair AI. We have noticed first-hand the prevalent role of CS in the produced outcomes of the project, thus additionally motivating our interest in how and to what degree CS researchers integrate the demands of SOI into their everyday research practices.

\paragraph{\textbf{Contribution.}} 
The objectives of this paper are threefold. First, from a conceptual angle, we seek to reintroduce the longstanding and classical approach of articulation work by \citet{fujimura_constructing_1987}, inspired by the work of \citet{corbin_articulation_1993}, into CS and CSCW. Despite Fujimura's significant contributions to the social studies of science\footnote{{\href{https://4sonline.org/2023_joan_fujimura_and_warwick_anderson.php}{Bernal Prize 2023: Joan Fujimura and Warwick Anderson. Society for Social Studies of Science}.}} and the concept's past significance for CSCW \cite{schmidt_taking_1992}, we have noticed that it has somewhat faded from contemporary discussions. Notwithstanding, we believe it is still an important lens for CSCW and CS today because it enables a closer examination of research and its (shifting) constraints in times of the increasing uncertainty of academic labor \cite{sigl_tacit_2016}. It is also important because it foregrounds that mundane and tacit organizational efforts are always necessary if novel technologies (such as fair AI techniques) are to become integrated into already situated practices and organizational settings \cite{grankvist_making_2011}.

The concept of articulation work also applies to other areas of CS research, as well as to research in general. However, studying the case of fair AI is particularly interesting given the increasing role of social world actors, such as regulatory and policy bodies. Our second contribution is thus to provide a deeper understanding of fair AI research as the emergent product of effectively aligning one's research with social world actors. We have personally experienced the effects that these actors can have on laboratory practices, and we consider it crucial to thoroughly examine the influence of such interactions. 

Third, we are acutely aware that all too often, there is far too little time for a meaningful reflection of our daily work. The insight that our research is contingent on what is ``doable'' may initially seem quite straightforward. However, it also raises the question of how we can potentially expand the scope of doability. Connecting this to the identified gap between CS and SOI, we thirdly contribute with a reflection on how SOI can become a more doable problem.

\paragraph{\textbf{Main findings.}}
We first start with a presentation of the most important social world actors and their (partially ambiguous) demands. We find that there is an increased orientation towards the SOI paradigm for most actors, but also a lack of clarity about what \textit{doing} SOI research means.
We then zoom in on the everyday practices on the level of the CS laboratory. We show that interview participants engage in SOI to some extent, but everyday work (e.g. writing proposals, securing funding and staff, reading and producing papers), technical requirements (e.g. obtaining data and models), and temporal constraints challenge alignment with SOI, and restrict their ability to work beyond disciplinary expectations. We conclude our findings with the micro-strategies of the interview participants to integrate SOI into their everyday research practices.

\paragraph{\textbf{Structure of the paper.}}

The rest of this paper is organized as follows: 
In Section \ref{sec:background}, we give a brief overview of the evolvement of the two paradigms, CS and SOI, along with the development of fair AI research. We further outline important related work on the organizational challenges and constraints for fair AI development and research.
In Section \ref{sec:case}, we introduce the concept of articulation work, and explain how the doability of a research problem is contingent on the successful alignment across the three levels of work organization.
In Section \ref{sec:method}, we present our qualitative study and offer insights into the selection, academic background, and scientific impact of our interview participants. We explain our approach to qualitative data coding and provide a detailed account of how we follow ethical standards in our interview study.
In Section \ref{sec:findings}, we present a comprehensive analysis of our research outcomes.
In Section \ref{sec:discussion}, we discuss the role of facilitators and constraints in shaping the current state of fair AI research and its implications for making SOI-oriented fair AI research doable. We address the limitations of our research and propose future work in Section \ref{sec:limitations}. Finally, Section \ref{sec:conclusion} gives a brief reflection on potential directions for fair AI research.

%% file: sections/02background.tex
\section{Background and related work}
\label{sec:background}

Algorithmic-decision systems that draw on AI and machine learning (ML) techniques are proposed and deployed for a myriad of tasks across domains, some with high-stakes implications, e.g., healthcare treatment allocation, credit assessment, or job suitability. Their implementation is supported in great part by the promise ``to bring greater discipline to decision-making'' \cite{barocas-hardt-narayanan}. However, alongside the potential societal advantages attributed to the use of these systems, their deployment presents a challenge to society: they can perpetuate, and render invisible, structural forms of discrimination, such as sexism, racism, classism, and ageism. Extensively researched instances of algorithmic discrimination \cite{angwin_machine_2018, buolamwini_gender_2018, lopezBiasDoesNot2021, miceli_studying_2022} and media coverage of cases of algorithmic bias\footnote{{\href{https://www.taylorfrancis.com/chapters/edit/10.1201/9781003278290-44/amazon-scraps-secret-ai-recruiting-tool-showed-bias-women-jeffrey-dastin}{Dastin 2018. Amazon Scraps Secret AI Recruiting Tool that Showed Bias against Women. Reuters.}}} \footnote{{\href{https://eu.usatoday.com/story/tech/2015/07/01/google-apologizes-after-photos-identify-black-people-as-gorillas/29567465/}{Guynn 2015. Google Photos labeled black people 'gorillas'. USA TODAY.}}} \footnote{{\href{https://www.propublica.org/article/machine-bias-risk-assessments-in-criminal-sentencing}{Angwin and Larson 2016. Machine Bias. ProPublica.}}} have highlighted the imperative to detect and mitigate such harms. This has been paralleled by a growth in the number of publications about AI ethical principles and guidelines \cite{jobin_global_2019}, the proposal of AI regulation across the globe \cite{european_commission_proposal_2021,office_of_science_and_technology_policy_blueprint_2022,department_of_industry_science_and_resources_artificial_2023,yang_china_2023}, and the flourishing of related research communities across all disciplines. In this specific conjuncture, the field of fair AI has also expanded across differing AI and computational sub-fields.

Fair AI is currently situated at an ``intermediate stage of disciplinary evolvement'' \cite{heilbron_regime_2004, merz_local_2016}. In this stage, there is a growing importance of SOI research, placing CS research on fair AI more and more at the intersection of these two research paradigms. As \citet{benbouzid_fairness_2023} notes in regard to the increasing SOI demands for transparency, CS researchers ``face two difficult and often distinct types of demands: first, for reliable computational techniques, and second, for transparency, given the constructed, politically situated nature of quantification operations''. 

Transparency is a characteristic that is asked of fair AI research, however within our context, it is only one of the many indicators of the sociopolitical dimensions of this type of research. To expand on our introductory definition, we understand and use SOI as an umbrella term encompassing a broad range of research topics. SOI involves social scientist researchers as well as various societal actors, such as regulatory bodies, policymakers, ethicists, advocacy groups, and the broader community affected by AI technologies \cite{barocas-hardt-narayanan}. The gap between SOI and CS for fair AI also has led to the creation of, and changes in, already existing publication venues and conferences. The first Association for Computing Machinery\footnote{\href{https://www.acm.org/}{Association for Computing Machinery (ACM) landing page}}(ACM)-sponsored conference organized by the FAT* network, now known as ACM Conference on Fairness, Accountability and Transparency (ACM FAccT), took place in 2018, leading to it becoming a renowned academic venue for innovative research on fairness, accountability, and transparency in ML with an interdisciplinary orientation\footnote{{\href{https://youtu.be/g3drDUAw2jE?t=1596}{ACM FAccT Conference: FAccT'22 Town Hall Meeting (\textit{accessed via YouTube, 2024}).}}}. The conference has striven to feature impactful work, and has given visibility to the ongoing criticism geared towards AI, such as the pursuit of technical progress at great material and environmental costs \cite{10.1145/3442188.3445922}; inherent functional and external validity issues \cite{raji_fallacy_2022}, and going further than that, its research values and foundations, as the work by \citet{birhane_values_2022} has shown. 

Alongside ACM FAccT, recent years have seen the emergence and establishment of other SOI-inclusive venues, such as AAAI/ACM Conference on AI, Ethics, and Society (AIES), and ACM conference on Equity and Access in Algorithms, Mechanisms, and Optimization (EAAMO); these venues increasingly encourage interdisciplinary work, and are helmed by seemingly interdisciplinary leadership.
Nevertheless, the gap between paradigms remains, and it can be seen in how the CS research community relates to and pursues different publishing venues. Given the popularity and interest of the fair AI topic, the last years have allowed for the emergence of new venues in the CS paradigm that explicitly seek to form a community supportive of primarily computationally rigorous research applied to ``problems of pressing and anticipated societal concern'' (e.g., Symposium on the Foundations of Responsible Computing). Similarly, there is now also the openness of traditional CS venues such as NeurIPS, the International Conference on Machine Learning (ICML), or the European Conference on Machine Learning and Principles and Practice of Knowledge Discovery in Databases (ECML-PKDD) to computational research on fair AI, via the main tracks, specially created ones, or thematic workshops. A more recent emerging trend we have also observed from leading and traditional CS venues (e.g., European Conference on Information Retrieval, and the  International Joint Conference on Artificial  Intelligence) is the creation of special research tracks for AI that address societal concerns dedicated to work carried out in active collaboration with civil society stakeholders. Thus, when we refer to CS researchers aligning with SOI, we mean that they either engage in epistemic research topics that concern the SOI paradigms, engage in collaborations with SOI actors, or publish in SOI venues.

Studies of the current tensions within fair AI have identified long-standing standards of practices, decision-making power imbalances, and fuzzy best practice guidelines to be all persistent challenges  \cite{widder_limits_2022,orr_attributions_2020}. \citet{sloane_german_2022} also emphasize how ML systems, AI ethics, and social practices cannot be disentangled from their specific cultural and historical contexts. 
Multiple studies, \cite{holstein_improving_2019,veale_fairness_2018,department_of_industry_science_and_resources_artificial_2023,heger_understanding_2022}, explore the gaps between research and the mundane, organizational work of CS researchers. Particularly, \citet{holstein_improving_2019} and \citet{veale_fairness_2018}, address existing challenges faced by practitioners in the private and public sectors to develop and incorporate fairness notions in machine learning systems, while \citet{deng_exploring_2022} and \citet{heger_understanding_2022} examine the particulars of integrating fairness toolkits and documentation frameworks within organizational workflows.

Conclusions drawn from these studies coincide with the impossibility of a frictionless integration of technical interventions, despite the attempts by computer scientists \cite{holstein_improving_2019}. The literature also attributes rising frictions in the process of implementing values in algorithms to other organizational issues such as time constraints, business objectives, and skill mix issues. We set out to contribute to these studies by introducing the concept of articulation work and using it to unravel further the tensions of fair AI research as a problem of organizational alignment.

%% file: sections/03case.tex
\section{Articulation work and the doability of research}
\label{sec:case}

\begin{figure}[!t]
    \centering
\includegraphics[width=.99\linewidth]{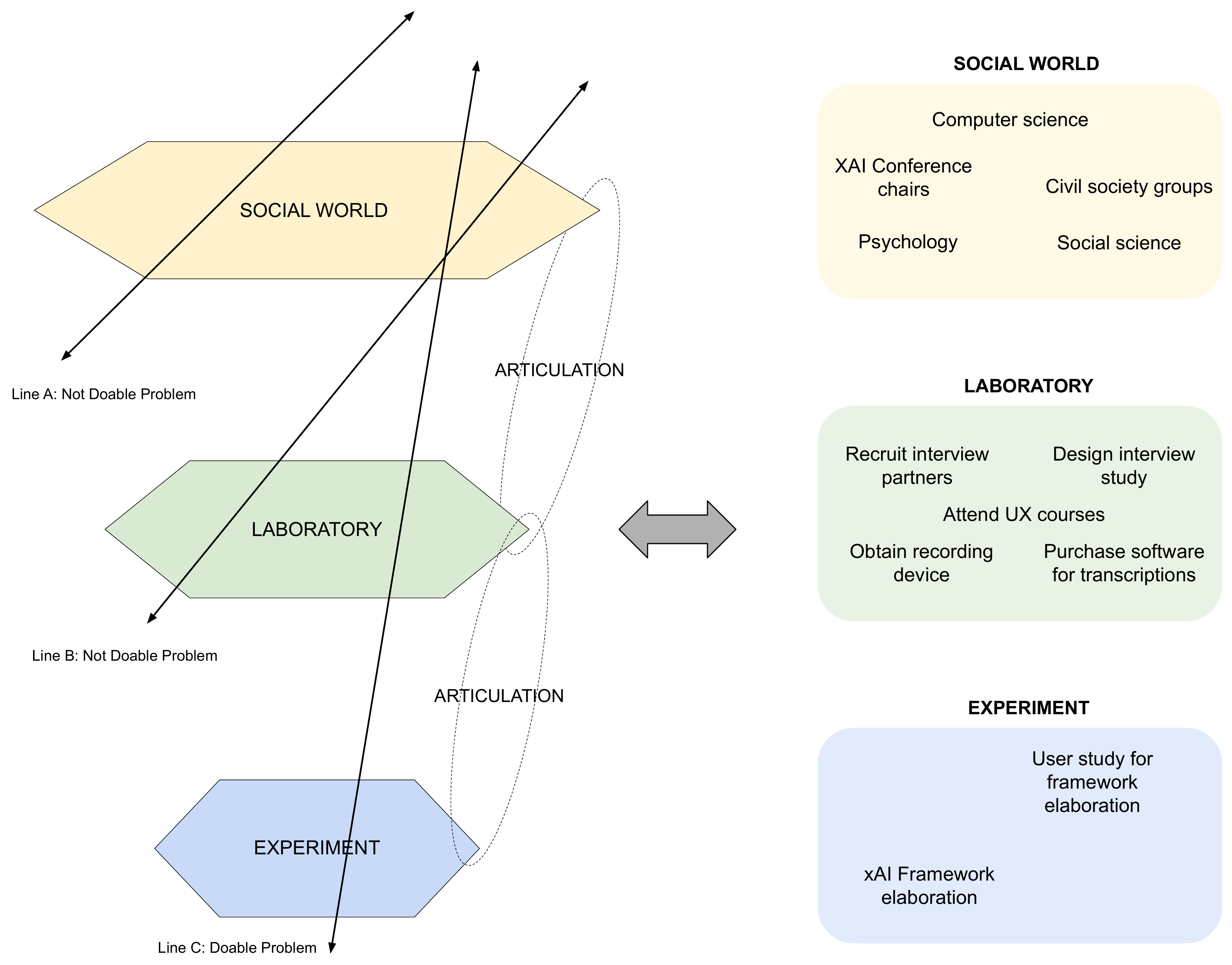}
\Description[Visualization of the levels of work organization represented with three polygons of different colors, side by side with a practical example.]{On the left side of the illustration, there are three polygons organized vertically. At the top is the largest one, yellow-colored. It represents the Social World. The polygon in the middle is shorter, and green-colored, representing Laboratory Level. The one at the bottom, the smallest, and blue-colored, represents the Experiment Level. Here we have also three arrows illustrated, which represent research alignment for doable research problems. The first arrow (a) that only goes through the Social World is not a doable problem. The second arrow (b), which goes through the Social World and the Laboratory, also denotes a problem that is not doable. The third arrow, (c), goes through all three levels and thus denotes a doable problem. Further, we also include two ovals, one connects the Social World with the Laboratory to denote articulation work in between levels, and another one between Laboratory and Experiment. A bidirectional gray arrow unites this with the right side of the image. Here we have three-colored rectangles with rounded corners that contain details of our examples. In the Social World rectangle, which is yellow, we include the words: computer science, xAI conference chairs, psychology, and social science. In the Laboratory rectangle, which is green, the words are: Recruit interview partners, design interview study, attend UX courses, obtain recording device, purchase software for transcriptions. For the Experiment rectangle, which is blue, the words are: User study for framework elaboration, xAI framework elaboration.}
\caption{\textbf{Fujimura's three levels of work organization \cite{fujimura_constructing_1987}, adapted for the case of fair AI.} On the left side of the illustration, the three levels of work organization as per Fujimura's \cite{fujimura_constructing_1987} original conceptualization are depicted. To enable alignment, articulation work is needed, between levels. Further, the labeled lines (A \& B) show how without alignment across all levels, doablity is not possible. On the right, we identify examples of actors and tasks that can be found across the three levels in line with our example.}
\label{fig:doability}
\end{figure}

Science encompasses more than knowledge, it involves work. This basic insight stems from research in CSCW and STS, that emerged in the early 1970s as a new way of thinking about science \cite{pickering_science_1992}. There are two distinct forms of scientific work. The first, known as \emph{production work}, involves translating specific research interests into concrete research outcomes. This involves applying scientific methods, conducting experiments, analyzing data, and, in a general sense, engaging in thinking \cite{fujimura_constructing_1987, knorr_tinkering_1979}. On the other hand, there is a second type of work conceptualized as \emph{articulation work}, also described as ``work to make work work'' \cite{pallesen_articulation_2018, schmidt_remarks_2011}. Articulation work involves daily tasks like planning, organizing, monitoring, evaluating, adjusting, coordinating, and integrating activities. 

According to \citet{fujimura_constructing_1987}'s ethnographic research in the field of biomedical cancer research, articulation work is crucial for rendering a scientific problem \emph{doable}. Doability is achieved by establishing \emph{alignment} across three levels: the experiment, the laboratory, and the social world \cite{corbin_articulation_1993, strauss_work_1985}. On a micro level, the \emph{experiment} encompasses a set of tasks conducted within the laboratory. The \emph{laboratory} serves as the physical space for experiments and related activities. Expanding to the macro level, the \emph{social world} is the broader context where experiments and laboratories are situated. The social world is the arena where all collective and individual actors involved in a specific research problem interact. A social world is characterized by specific activities, sites and requirements, and existing or new technologies that enable the fulfillment of the social world’s requirements \cite{clarke_social_2008, strauss_negotiations_1978}.

To provide a clear understanding of these abstract concepts, let us imagine a PhD student in a computer science laboratory (CS lab). The PhD student wants to validate their explainable ML (xAI) framework with a user study in collaboration with a principal investigator (PI) who is an expert in user research. The student successfully gathers several factors that are situated at the experiment and laboratory level, e.g., attending user experience design courses, designing the study, recruiting interview participants, and obtaining a recording device and software for transcriptions. In alignment with the social world, the student uses a common explainability technique acknowledged by the CS community. The CS community also supports user studies as a valid method for assessing the interpretability of explanations. Further, the student's research connects to other academic disciplines such as social sciences and psychology, and civil society groups' concerns on opaque algorithms. The student's efforts to combine all these factors, and connect them to the social world's requirements, is thus what makes their research doable. Over time, the student's research objectives can also change to maintain alignment \cite{grankvist_making_2011}. Figure \ref{fig:doability} illustrates Fujimura’s conceptualization of the organization of work across the three different levels (social world, laboratory, experiment) on the left-hand side, and our example divided into these levels on the right-hand side.

\paragraph{\textbf{Facilitators and constraints.}} \citet{fujimura_constructing_1987} examines three conditions of scientific research that facilitate or decrease articulation work. The \textit{leeway provided by available resources}, and a \textit{clear division of labor} facilitate articulating alignment. Conversely, uncertainty limits researchers' ability to plan. Consequently, more articulation of work is then needed to conduct research spontaneously. As for the PhD student from our example, constraints to performing their xAI user study could be the unavailability of courses during the semester, or the unwillingness of potential interview participants to remain throughout the length of the study. The student also needs to complete a convincing research proposal to get the support of their supervisor, regularly correspond with the hosting PI, and engage and survey relevant literature to support their findings. However, in the event the hosting PI does not see value in performing the study, or the supervisor faces time constraints, articulation work for the student will increase significantly. If alignment with important social world actors is not possible, the research will not be doable.

\paragraph{\textbf{Tinkering.}} Constraints and uncertainties evoke tinkering, whereby scientists cope with difficult situations and unexpected problems \cite{lin_bias_2023}. Tinkering is an idiosyncratic, situated scientific practice in which local opportunities are made to work to solve a problem \cite{knorr_tinkering_1979, nutch_gadgets_1996}. Considering the scenario in which the supervisor initially hesitates due to time constraints, the PhD student might tactfully adjust the research timeline or propose more flexible arrangements. The student might also suggest using video updates to maintain engagement without the need for simultaneous, time-intensive meetings with the supervisor. Such small adjustments exemplify how tinkering can introduce creative solutions to unforeseen challenges.

To summarize, the doability of a research problem is contingent on the successful alignment across three levels of work organization, and shaped by facilitators and constraints, scientists’ actions and positions \cite{larregue_long_2018, haraway_situated_1988, harding_whose_2016}. If an unexpected problem arises, scientists engage in tinkering to maintain alignment with social world actors. Nevertheless, the impacts of organizational conditions on research are often overlooked, neglected, or obscured \cite{lampland_standards_2009, nissenbaum_how_2001}.

%% file: sections/04method.tex
\section{Method and Data Analysis}
\label{sec:method}

\citet{selbst_fairness_2019} identify CS researchers working in fair AI as powerful. This is because the form a technology ultimately takes is shaped by the perspectives and practices of those that develop it: ``as fair ML researchers seek to define the `best' approach to fairness, we also implicitly decide which problems and relevant social groups are important to include in this process. Our choices prioritize certain views over others, exerting power in ways that must be accounted for'' \cite{selbst_fairness_2019}. Not least for this reason, we study the gap between SOI and CS, starting from the influential perspectives and practices of CS researchers.

\paragraph{\textbf{Data collection and interview partners.}} Our qualitative study was completed within the context of an EU-funded project, dedicated to researching and developing interdisciplinary methods for fair AI systems. In addition to this particular affiliation, our interview partners were selected following these concise criteria: (i) they held a PhD in computer science, (ii) they held the role of PI in at least one research project, (iii) they worked on research topics about fair AI to varying degrees of involvement, (iv) they were based in different CS laboratories across different university departments or research centers. Our initial list was made up of twenty prospective participants, all identified through the directory of the project network. Over five months, we progressively contacted all of them via e-mail, introducing them to the general idea of the study, the methodology, and disclosing possible publication purposes. 

To our interview requests, we received eleven positive answers, with ten following through with the interview. The resulting group was therefore made up of established, fairly senior, CS researchers, who are currently actively involved in fair AI research in Europe. Table \ref{tab:interviews} displays relevant information about our interview partners, such as the institutional position, country of institution, and gender. 

Even though almost half of our cohort indicated they had been involved in fair AI research for a long time, some of our interview partners were relatively new to the field and even mentioned that fair AI research was not their primary area of expertise. In the latter instance, their CS sub-fields of principal research included computational social science, social network analysis, information retrieval, semantic web, or data mining, to name a few. 

\begin{figure}[t]
    \centering
\includegraphics[width=90mm,height=60mm,keepaspectratio, trim = 0.5cm 0.5cm 0cm 0.5cm, clip]{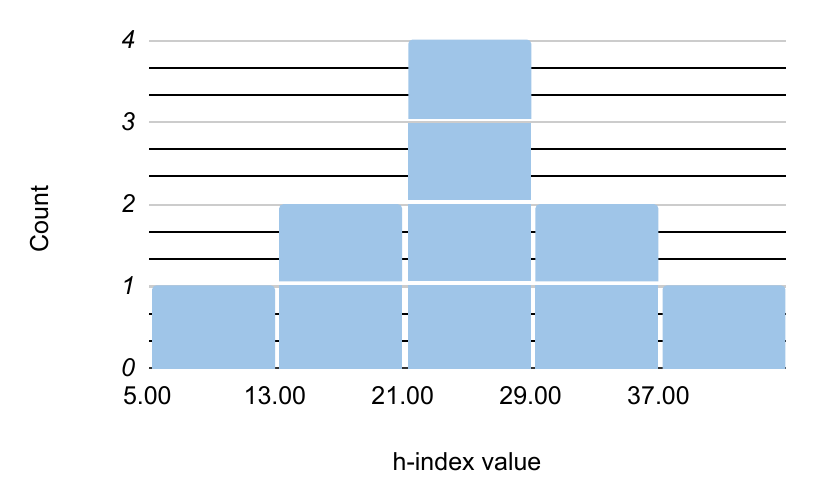}
\Description[A histogram was used to show the distribution of the h-index of the study.]{Depicted is a histogram in blue colored boxes, with the y-axis showing the count, from 0 to 4 and the x-axis the value of the h-index, 5, 13, 21, 29, 37. There is one individual with an h-index between 5 and 13, 2 individuals with an h-index ranging from 13 and 21, 4 individuals with a value between 21 and 29, 2 individuals with an h-index value between 20 and 37, and one individual with an h-index value higher than 37.}\caption{\textbf{Histogram Distribution of the h-index* for all 10 participants (*last five years).}}
        \label{fig:h-index}
\end{figure}

\begin{figure}[t]
    \centering
  
\includegraphics[width=0.99\linewidth]{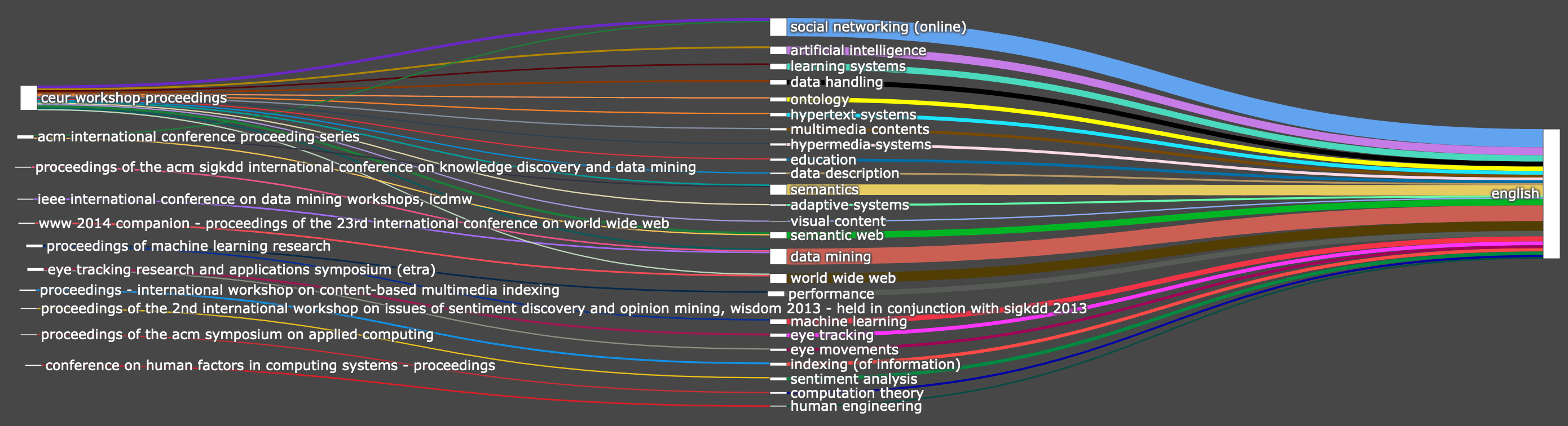}
\Description[A sankey diagram was used to give an overview of the publication venues and research sub-areas where our interview partners participate (last five years).]{A sankey diagram illustrates the relationships between publishing venue and research sub-area based on 450 research papers belonging to the 10 interview partners. On the left side of the diagram, there are 11 names for conferences and workshops. From top to bottom: CEUR workshop proceedings, ACM international conference proceeding series, proceedings of the ACM SIGKDD conference, IEEEE international conference on data mining, International conference on World Wide Web, proceedings of machine learning research, eye tracking research and applications symposium, content-based multimedia indexing, issues on sentiment discovery and opinion mining, ACM symposium on applied computing, conference on human factors in computing systems. This site is then linked to the middle column containing the research sub-area by colored, curved lines. From the top: social networking, artificial intelligence, learning systems, data handling, ontology, hypertext systems, multimedia contents, hypermedia systems, education, data description, semantics, adaptive systems, visual content, semantic web, data mining, world wide web, performance, machine learning, eye tracking, eye movements, indexing of information, sentiment analysis, computation theory, human engineering. These items are all joined with the column on the right, which represents the language of the published works (English). They are also joined with curved, colored lines that differ in thickness depending on the number of research papers.}\caption{\textbf{Sankey Diagram.} Represented are 450 conference and workshop papers from 2009 to 2023 from our interview partners to give an overview of publication venues and research sub-areas.}
            \label{fig:publishedworks}
\end{figure}

To further characterize the professional trajectory and disciplinary alignment of our interview partners, we performed a quantitative analysis of their scientific research. For this, we looked at a measure for scientific research output (i.e., h-index\footnote{reported as per user profiles on Google Scholar: \url{https://scholar.google.com/intl/en/scholar/metrics.html}} \cite{Hirsch2009AnIT}) and performed an assessment of their scholarly literature in a broad sense. Regarding the inclusion and utility of the h-index, we acknowledge both the limitations and the criticism directed towards such measures. For that reason, our objective is only to  provide an overview of the cumulative impact and perceived relevance of the scientific research output for all our interview partners in the last five years via a generally acknowledged indicator in academia. In Figure \ref{fig:h-index} we display the distribution of this index for our participants.  
In Figure \ref{fig:publishedworks}, we also present a Sankey Diagram generated by executing the Python library pyBibX\footnote{\url{https://pypi.org/project/pyBibX/}}. This visualization enables us to cumulatively analyze raw data files representing the individual author's profiles for all ten of our interview partners, sourced from the scientific database Scopus\footnote{\url{https://www.scopus.com/home.uri}}. Concisely, the diagram illustrates the flow of 450 published works by our interview partners for the years 2009-2023, between the top 20 conferences and workshops the papers were accepted at and the research sub-areas the papers belong to. In addition to capturing this document flow, the analysis also makes it possible to denote how CS-oriented publication venues (e.g., ACM International Conference on Knowledge Discovery and Data Mining, IEEE International Conference on Data Mining, The Web Conference, ACM Symposium on Applied Computing) and research sub-areas (e.g., online social networking,  semantics, eye tracking, sentiment analysis) are predominantly chosen by our interview partners.

\begin{table}[!t]
\centering
\caption{\textbf{Demographic information and citation metrics of interview partners}}
\label{tab:interviews}
\resizebox{0.6\columnwidth}{!}{%
\begin{tabular}{|c|c|c|c|c|}
\hline
\rowcolor{pink!40} 
\textbf{Identifier} & \textbf{\begin{tabular}[c]{@{}c@{}}Institutional\\ Position\end{tabular}} & \textbf{\begin{tabular}[c]{@{}c@{}}Country of\\ Institution\end{tabular}} & \begin{tabular}[c]{@{}c@{}}\textbf{Gender}\end{tabular} \\ \hline
R1 & Senior Researcher & Germany & M  \\ \hline
R2 & Senior Researcher & Greece & M  \\ \hline
R3 & Full Professor & Germany & M  \\ \hline
R4 & Full Professor & Germany & F  \\ \hline
R5 & Full Professor & Germany & F  \\ \hline
R6 & Full Professor & United Kingdom & M  \\ \hline
R7 & Full Professor & Netherlands & M  \\ \hline
R8 & Full Professor & Italy & M  \\ \hline
R9 & Full Professor & Belgium & M  \\ \hline
R10 & Associate Professor & United Kingdom & M  \\ \hline
\end{tabular}%
}
\end{table}

\paragraph{\textbf{Interview process and qualitative data coding.}} 
The completed qualitative study follows an episodic narrative interview approach \cite{mueller_episodic_2019}. This method is used to improve the understanding of a phenomenon, in our case fair AI, by inviting the interview partners to tell their individual stories of life and work experience around said phenomenon. The interview format allows interview partners enough time for self-reflection and for the opportunity to share in richer detail their direct experiences. The collected data is thus useful to uncover the narratives around fair AI and related activities, which then require analysis and interpretation \cite{silverman_doing_2010}. 
Our interviews were held throughout the year 2022, with the duration of each interview ranging from one to one and a half hours. Following every interview, the recordings were manually transcribed and pseudonymized. All personal identifiers were removed from our interview data and all our interview partners are referred to by pseudonyms i.e., R1, R2, etc. 
Subsequently, the interview material was interpreted using an explorative approach, informed by an inductive analysis that facilitates the possibilities of finding novel and unexpected issues \cite{corbin_grounded_1990}. The first iterative phase of interpretation had been conducted closely to the material, resulting in hundreds of qualitative codes. Following this, we conducted the rest of the coding process, gradually clustering the codes into certain categories, merging, and rephrasing codes in numerous iterative phases. The resulting codes and sub-codes were reconsidered and collectively discussed many times.

\paragraph{\textbf{Research Ethics.}}
All participants in our study were provided with and acknowledged a consent form before the interviews. The document outlined the objectives of our project, the data collection process, transcription, and analysis procedures. It explicitly highlighted voluntary participation, emphasizing participants' right to terminate interviews at any point (reiterated before the interview). We conducted all interviews through an open-source video conferencing platform hosted on the server of one of our academic institutions. Furthermore, all data was stored on servers located in Europe, aligning with the regulations of the EU General Data Protection Regulation (GDPR) legislation.

%% file: sections/05findings.tex
\section{Findings}
\label{sec:findings}
Our findings are structured into three subsections. We start with two questions: \textit{First,} \textit{what does the fair AI world want?} Second, zoomed in on the CS lab, \textit{what do CS researchers do?} Of course, the fair AI world also `does things', and CS researchers also `want things'. Yet in this particular case, our focus lies on the (SOI) demands of the fair AI world to understand whether and how they align with what is done in the situated practice of the CS lab. In the last subsection, we further zoom into researchers' micro-strategies for including and tinkering with SOI demands.

\subsection{Social worlds: what does the fair AI world want?}
\label{aligning:social_world}
In this section, we introduce the most prominent actors in the social world of fair AI that we identified in our interviews. We refer to it herein as the fair AI world. Our first finding is that there are three important collective actors: the CS community, project partners, and regulatory bodies. These actors display an increasing orientation towards the SOI paradigm but with different needs, expectations, and interpretations attached to it.

\paragraph{\textbf{CS community.}} 
The first significant actor that emerged in the interviews was the CS community. However, there was a time when our early contributors to fair AI found it challenging to align their research with the then-research interests of the CS community. Particularly, the CS community was mainly concerned with research problems that tackled scalability or efficiency, as retold by R8, a computer science professor at an Italian university. It was, especially before 2015, that `fairness' was neither a prominent research area in the CS community nor the widely popular topic it is today. R1, a senior scientist working at a German research lab, recalled that time as follows:
{\vspace{1mm}
\textit{\begin{quote}
``I googled fair analytics and the only thing back then which I found was [...] a job fair where people could look for analytics jobs. You could find nothing on the internet about fair analytics.'' (R1). 
\end{quote}}
\vspace{1mm}}

Another professor at a Dutch university told us that he needed to convince his colleagues about fairness being an important and relevant research problem for the CS community.
\vspace{1mm}
\textit{\begin{quote}
``It was a relatively tough period to convince research communities about this research we are doing because most of the computer science, machine learning AI world did not recognize it as important for the first few years.'' (R7)
\end{quote}}
\vspace{1mm}

Interests and demands of social worlds, such as in the case of the CS community, can shift and change \cite{knorr-cetina_epistemic_1991, knorr-cetina_epistemic_1999}. Once the CS community acknowledged research on fairness (metrics), the initial misalignment with the CS community ``\emph{does not hurt anymore}”, to further quote R7. Early contributors also outlined how they played a crucial role in shaping the community's evolving interests in fairness. Notably, the impact of their contributions is evident, if only taking into account citation figures. For instance, R9's first fairness paper has over 1000 citations, while an early contribution by R8 surpasses 4000 citations, at the time of writing. There is a successful alignment of CS researchers' work on fairness with the demands of the CS community, and their impact increases as the fair AI field grows.

\paragraph{\textbf{Project partners.}} 
The second significant actor was the `project partner'. Typically, we found a project partner to be a private entity, and in our interviews, banks, credit agencies, and insurance companies were the ones most frequently mentioned. Public project partners were institutions such as hospitals, police, public broadcasters, or other research institutions. While there are many challenges for researchers in applied partnerships due to, for example, privacy concerns or licensing disputes \cite{holloway_normalizing_2015, biscotti_constructing_2012}, private and public actors were perceived as particularly valuable because they can be a source of funding, interesting use cases, and datasets and models to study fairness. The latter was understood to be important for fair AI research because it required societal datasets containing information about people. A successful alignment with such data-providing partners was termed `real-world' research, and demonstrating real-world impact was considered to be exceptionally relevant by our interview participants. 
\vspace{1mm}
\textit{\begin{quote}
``I will talk about this project because it is one of the latest and it’s still running and it is a real project, so I’m proud, compared to projects which are theoretical, lab-based.'' (R8)
\end{quote}}
\vspace{1mm}
Our interview participants attentively observed their project partners' (and, in alignment with this, also their experiments and CS laboratories') increasing orientation towards fairness and ethics in general, with R2 describing it as a paradigm shift. 
\vspace{1mm}
\textit{\begin{quote}
``I actually see this as a kind of paradigm shift. Several years back when I was a young researcher, ethical aspects in our kind of research were kind of an afterthought and a necessary evil but right now it’s kind of the first thing to consider and it’s always there'' (R2). 
\end{quote}}
\vspace{1mm}
Other interview participants expressed a more negative perspective, characterizing an orientation towards fairness as a competitive advantage for institutions, that allowed them to strategically position themselves as socially responsible and attentive to the societal impact of their AI systems. They had witnessed instances where project partners guided and constrained research directions and objectives. For example, R8 told us that in a joint research project with a bank, the development of a fair credit scoring model was curtailed due to monetary restrictions.
\vspace{1mm}
\textit{\begin{quote}
``In theory, we could have developed different models, one predicting the default within six months, one within one year, and so on, so that every user could use the model for the period preferred. But this was cost-intense, e.g. building, maintaining several models, and so in the end it was not possible for the budget of the company to do that.'' (R8) \end{quote}}
\vspace{1mm}

Private project partners also seemed to provide a (limited) playground for SOI research, while CS research remained the primary objective \cite{10.1162/qss_a_00267}. We observed that CS researchers sometimes performed as qualitative researchers. In one case, a professor encouraged students to carry out ``short ethnographic studies'' at the institutions of their project partners. We interpret this practice as a form of tinkering and provide further details in Subsection \ref{aligning:tinkering}. However, interview participants ultimately had to align their research with the requirements of project partners, and the distribution of power between researchers and partners was not always balanced.

\paragraph{\textbf{Regulatory bodies.}}
Regulatory bodies, including various institutions within the EU, emerged as the third important social world actor for our interview participants. Not only did they serve as (novel) funding bodies for fair AI research, but it was expected that they would also progressively be taking on roles as project partners themselves.
Across the globe, proposals for fair and responsible AI are being introduced and debated, with the European Union’s AI Act being the first of its kind with regulatory enforceability in the EU \cite{european_commission_proposal_2021, office_of_science_and_technology_policy_blueprint_2022, department_of_industry_science_and_resources_artificial_2023, yang_china_2023}.

The majority of our interview partners anticipated advantages for their research once such regulations were implemented. For instance, R1 emphasized the necessity for \textit{``some common ground where you can stand on, and these are standards''}. In a similar vein, R8 suggested standardization bodies like the International Organization for Standardization (ISO) control and audit AI development. He also proposed compensation mechanisms for \textit{``people who have been harmed by such systems''}, and the introduction of general reporting mechanisms. These reporting systems would, as R8 explained, enable potentially disadvantaged individuals to denounce and record their cases.

Despite this predominantly expressed desire for standards and reporting mechanisms, at the time of the interviews, our participants mainly engaged with regulatory bodies through the perspective of their project partners. R8 noted how current regulatory attempts shifted his project partners' interest toward identifying and mitigating algorithmic bias. He had observed a similar pattern in the past regarding (computational) privacy issues, wherein significant interest in the topic only emerged after the implementation of regulations.
\vspace{1mm}
\textit{\begin{quote}
``I have to say that the [company's] interest in these [fairness] techniques was motivated by the rules, by the law. This is a general behavior, for instance, [if you] think about privacy twenty years ago, nobody cared about privacy issues until the regulations came out.'' (R8)
\end{quote}}
\vspace{1mm}
A similar viewpoint was shared by R7, regarding a project with an insurance company to build risk models and implement fairness techniques. 
\vspace{1mm}
\textit{\begin{quote}
``Now, their interest is, as far as I can see, not per se because they want to make what is currently there, fair. They do not necessarily consider it unfair, let’s put it that way. But they want to be prepared and even steer the discussion if at a certain point legislation may come that will force them to become, to obey some fairness measures.'' (R7)
\end{quote}}
\vspace{1mm}

We can see that in this ambivalent period, social world actors were connected, showing different interests in influencing each other. If project partners' interest in fairness were mainly driven by business needs, there was the possibility that they might be simply interested in achieving basic requirements for legal compliance, or in more extreme cases, they might be interested in becoming a key player in defining upcoming regulation, as has been reported\footnote{\href{https://time.com/6288245/openai-eu-lobbying-ai-act/}{Perrigo 2023. OpenAI Lobbied the E.U. to Water Down AI Regulation. Times}}. Ultimately, neither of these scenarios could be perceived to be aligned with the expectations of the SOI paradigm.

\paragraph{\textbf{Summary.}} \textit{What does the fair AI world want?} We saw that the CS community's research acknowledged fair AI research as important, project partners were perceived to shape and constrain research problems, and regulatory bodies were emerging with open consequences for social world actors, and CS researchers. As a result, we found most interview participants to be uncertain about how to interpret current dynamics and in which direction the fair AI world would `truly' orient itself. As one interview participant, R1, slightly disillusioned, noted, \textit{``in the end, everyone has a different understanding of what is fair and what is not.''} Against this backdrop, the expressed desire for regulatory standards can also be considered as an attempt to obtain guidance on how fairness can become a doable problem. Yet, simultaneous alignment with multiple demands of diverse social world actors increased articulation work. This is because their perceptions, requirements, and demands were difficult to align.

\subsection{Laboratories and experiments: what do CS researchers do?}
\label{aligning:laboratory}
In this section, we zoom in on the level of the CS lab. We elaborate on the different tasks and factors that interview participants gathered to do fair AI research. We found that the organizational conditions of the lab steered researchers' attention towards other, more pragmatic concerns.

\paragraph{\textbf{Projects and staff.}} At the level of the lab, our interview participants were often and understandably occupied with more pressing tasks and concerns that did not necessarily pertain to tending to the broader social claims that motivated their research. They needed to spend most of their time dividing their tasks within and between projects, acquiring new projects, and strategically pursuing new projects with partners to gain access to new resources. For instance, R5 conveyed that her current research projects on fair AI also offered secure employment opportunities.
\vspace{1mm}
\textit{\begin{quote}
``To be honest, it's not really about the projects. Right? So the projects are just a funding tool, right?'' (R5)
\end{quote} }
\vspace{1mm}
In a similar vein, fair AI projects were also a way to secure positions for preceding and younger researchers, as R6 remarked that the fair AI research projects he was currently working on were also important because they secured his students' future positions in research. This is not to say that their research was not important to our interview participants, but, that on top of it, fair AI work had other dimensions beyond the undertaking of ethical and social justice-oriented claims. A fair AI laboratory is a workplace and also a site of career building for people. We witnessed that this reality manifested through practices such as placing staff on projects due to pre-existing working contracts or attempting to redefine work, to make it fit to existing knowledge and skills of the staff in the lab. 
\vspace{1mm}
\textit{\begin{quote}
``So you were sitting together with the configuration of people that is somehow given because they are there, and they are supposed to work now on something.'' (R4)
\end{quote}}
\vspace{1mm}

Some of our interviewed participants also saw how shouldering the responsibility to keep the staff in their labs employed created an unexpected incongruity. Here, we refer to the call to integrate diversity and inclusion (D\&I) considerations in AI systems and the claim that problems of unfairness and bias can be alleviated via diverse research teams \cite{Shams2023AIAT}. However, securing the mid to long-term positions of preexisting staff was seen to hinder these efforts. One interview participant was particularly concerned about this issue: 
\vspace{1mm}
\textit{\begin{quote}
``I think here we need to really think about how we can increase diversity, which is obviously important'' (R4).
\end{quote}}
\vspace{1mm}
It is clear that addressing the diversity of staff on projects required additional articulation work, and, if existing staff was not to be lost, additional resources were also needed to expand hiring possibilities.  

\paragraph{\textbf{Limited resources.}}
Project-based funding mechanisms also provided a temporary constraint to the distribution of resources, as project duration was always limited, typically spanning three to five years.
Such time constraints therefore called for savvy project management skills, particularly towards the end of a project life cycle. Inadvertently, this led to prioritizing research problems that contributed to achieving project objectives on time, but that perhaps required ``less effort'' from an experimental setup and execution point of view. For instance, although some interview participants attempted to set up user studies in their fair AI research, they expressed dissatisfaction as the actual engagement with users remained limited due to time constraints. R2 emphasized the typicality of such situations.
\vspace{1mm}
\textit{\begin{quote}
``I think this is quite typical for research projects that you don't get as far as to involve a real end user with your research software. Sometimes you manage to set up a kind of experimental setup but yeah I cannot think of any example that would have involved this [...] idea.'' (R2)
\end{quote}}
\vspace{1mm}
Some fair AI research problems, that were not deemed as required or necessary for successful project completion, were thus postponed, or not conducted at all.

\paragraph{\textbf{Noise and (other) papers.}}
According to our interview participants, the first step of the acquisition of a project involved the conception of an idea, followed by the publication of a corresponding paper. We witnessed that a published scientific paper also served as a legitimization of a novel research problem because it provided a first proof of doability. R6, among the well-experienced and influential researchers, described this legitimizing role of scientific papers as follows:
 \vspace{1mm}
 \textit{\begin{quote}
 ``I mean typically, the story is that you're interested in an area, you try to have maybe like a first initial paper and then you get funding.'' (R6)
 \end{quote}}
 \vspace{1mm}
Yet, our interview participants often grappled with keeping up with the overwhelming amount of new scientific papers and information that emerged in the fair AI research area. When we asked about their primary tasks, we heard that a significant amount of time was allotted to following, reviewing, and reading the numerous scientific papers being published, to remain informed and up-to-date. 
\vspace{1mm}
\textit{\begin{quote}
``Still, one big point is really to figure out what other people do on this topic. It's currently a hype topic and if it is a hype topic you have more publications than you can ever read.'' (R2)
\end{quote}}
\vspace{1mm}
Articulation work encompassed not only the task of monitoring and reading new papers, but also the discernment of which papers were merely contributing to such `noise', e.g. ``\emph{go[ing] on and on about the same things}” (R10), and those significant for fair AI research. In that regard, early-contributor R7 stated that \textit{``as soon as a particular few become hyped, over-hyped, many people jump on it and they aim to publish rather than to do something useful'' (R7)}. The days of these researchers being an `exception' for working on fairness research were long gone. In fact, due to the fast pace at which fair AI was evolving, CS researchers felt there were expectations to keep up with this high-speed publishing environment.
\vspace{1mm}
\textit{\begin{quote}
``I have to publish. I have to present the research results. Of course, we have to be productive. But I think it's good also to think a bit more, about what we want to do in the end. You can have better publications, [but] I think we [should] have better contributions, actually.''  (R5) 
\end{quote}}
\vspace{1mm}
Interview participants, such as R5, showcased that producing publications was a requirement for sustaining or enhancing academic careers. Becoming part of the `noise' was thus difficult to avoid, however, it also contributed to a lack of clarity about what constituted (or should constitute) impactful scientific progress for the fair AI world, even within the CS paradigm. 

\paragraph{\textbf{{Production work.}}} We now focus on the production work of fair AI research and related experiments. Early contributors recounted that their first experiments in fair AI research were conducted on an ad-hoc basis, such as training a classifier on a biased dataset to experiment with the results. After obtaining results, they would convince their colleagues (and themselves) that fairness was a relevant research problem by putting a valuable object in the foreground: the (biased) dataset.
\textit{\begin{quote}
``And it was one of those datasets that they used and that they showed that you could find [...] subgroups of people that were discriminated against.'' (R5)
\end{quote}}
\vspace{1mm}
Along with the dataset, other objects were placed in the foreground for production work. Over the past years, numerous fairness metrics and other bias mitigation techniques have been defined and developed, making it possible to ``package'' \cite{fujimura_constructing_1987} them into fairness toolkits \cite{10.1145/3600211.3604674}. These collections of metrics, methods, and practical instructions were intended to ease their use in research, but more importantly, streamline their use in industry, where often inexperienced practitioners were the ones tasked with building and `de-biasing' ML systems \cite{deng_exploring_2022,10.1145/3600211.3604674}. These objects thus facilitated alignment between the CS community and their project partners. And while 
the use of fairness metrics allowed for measuring, quantifying, and comparing the output of models, at the same time, they could instill the idea that only what can be captured by such metrics should be defined and identified as fair (or unfair).

R9, a professor at a university in Belgium and another early contributor to the field, emphasized that the primary objective of fair AI research should be to focus on model accuracy by \textit{``concentrat[ing] mainly on the distorted vision that is there [in the data used]''}. Accuracy, for him, was part of addressing fundamental computational and statistical questions that the CS lab was supposed to pose, and not ``\emph{political ones}”. With this demarcation approach, interview participants like R9 suggested that CS can and should be more autonomous from socially-oriented research and political actors. While such interpretations of the role of technology could be considered clear deviations from an SOI approach, we recalled that a focus on datasets and on computationally relevant questions such as accuracy had allowed for alignment with the initially dismissive CS community. In this way, fair AI research could now be ``stored'' \cite{star_ethnography_1999} (e.g. in a dataset, in a computer, in a metric), shared across different levels of work organization, and made comparable. Such practices of storing, sharing, and comparing, further facilitated alignment with other social world actors. For example, the creation of fairness metrics allowed for project partners to request that they be implemented into their own systems and provided a potential method to communicate standards adherence to regulatory bodies.

\paragraph{\textbf{Summary.}} \textit{What do CS researchers do?} In this section, we witnessed that CS researchers were engaged in a substantive amount of articulation work: writing project proposals, securing staff positions, allocating time, reading emerging papers, and coming up with new ones. All this deviated researchers' attention from engaging with fair AI research problems that they considered to be closer to the SOI paradigm. As a result, the production work of fair AI research was mainly aligned with the CS community and CS positions.

\subsection{Openness to the SOI paradigm and tinkering}
\label{aligning:tinkering}

We conclude our findings by pointing out interview participants' micro-strategies and attempts to integrate some SOI perspectives into their everyday research practices. 

\paragraph{\textbf{Openness to the SOI Paradigm}} Some of our interview participants, notably two professors, R4 and R5, both situated in German universities along with R8, the professor from an Italian university, and R10, a professor from a UK university, expressed their wish for fair AI research to become more inclusive of SOI perspectives. Some of them acknowledged their own agency and responsibility in attaining this. In particular, R10 emphasized how he tried to negotiate and shape the conditions of his CS lab to align with a more inclusive research approach. This, as he explained, included actively taking responsibility for conducting more research with interdisciplinary actors, such as scholars from the social sciences. As a senior researcher, he had experienced that such collaborations did not emerge spontaneously or by themselves but required proactive efforts by researchers, such as to \textit{``negotiate and have a more active role in your occupation to be able to keep the things that you're happy with.'' (R10)}

R5 invoked that the CS community must learn to perceive and do fair AI research differently than ``\textit{just another machine learning problem.}'' On the level of the lab and everyday research, this would include more reflection on fundamental questions in contrast to the more practical and technical aspects encountered in everyday tasks, for instance about the significance of the gender binary within datasets.
\vspace{1mm}
\textit{\begin{quote}
``Maybe thinking [of] fundamental questions again and kind of thinking from scratch or with a new perspective might be useful. I don't know how this perspective can come. I think looking at [fairness] as another machine learning problem is not a new perspective.'' (R5)
\end{quote}}
\vspace{1mm}

We further witnessed how several other researchers, such as R1 from the German research lab, R2 from the Greek university, and R7 from the Dutch university displayed some surface-level engagement with SOI actors in their research. They also called for greater inclusion of SOI actors, particularly, those people most affected (and potentially harmed) by a given AI system. In that regard, they all noted how ``the user''\footnote{The term user was also used to refer to employees in partner companies who may use an AI system in their decision-making process.}, as they described people at the receiving end of AI and ML models, had become more important. Alignment with the user happened mostly through their representation by civil society organizations in collaborative research projects. For example, R2 described one of his projects, where partners included government-affiliated (disabled) product user organizations. However, as previously detailed, user studies were still perceived as an additional task rather than an inherent step of CS research.

\paragraph{\textbf{Tinkering with SOI methods}} Some interview participants built their own accounts of doing SOI work by gradually deviating from usually performed CS tasks. Drawing on \citet{fujimura_constructing_1987} and \citet{knorr_tinkering_1979}, we understand these micro-strategies as \textit{tinkering}. 
We especially observed tinkering with SOI methods. For instance, R7, the professor at the Dutch University who, as mentioned in section \ref{aligning:social_world}, engaged his PhD students in conducting ``\textit{short ethnographic studies}'', as he coined them, at field sites of their collaborative partners. At the same time, the ethnographic study was further adapted and tinkered with, so that articulation work did not increase significantly. The professor conveyed that in his lab, ethnography - which is usually conducted for a longer period of time, with the researcher becoming very familiar with the field, following rigorous strategies of self-reflection, `thick description', and a back-and-forth between theory and empirical material \cite{hammersley_ethnography_2019, spradley_ethnographic_2016, geertz_thick_1973} - was only conducted for several days, and without formal training. In another example, R1 told us a story about how he aimed to conduct a qualitative study to test users' perceptions of an algorithmic system he and his colleagues were designing. However, to avoid additional articulation work (e.g. setting up a sampling strategy, reaching out to interview participants, organizing the interview setting, etc.), he turned to his colleagues as proxies for the intended users of the algorithmic system.

\paragraph{\textbf{Summary.}} Our findings show that there was a desire to accomplish impactful (fairness) research that coexisted with varying degrees of integration of SOI work into everyday research practices. Even researchers with a more critical stance towards the SOI paradigm demonstrated some (surface-level) engagement with it. Yet, due to the organizational constraints of the CS lab, SOI work was adjusted or modified, e.g. tinkered with, to sidestep additional articulation work.

%% file: sections/06discussion.tex
\section{Discussion}
\label{sec:discussion}
Bringing Fujimura’s study in dialogue with our findings, we now discuss the gap between the CS and SOI paradigms as a problem of ‘organizational alignment’. 

\paragraph{\textbf{Doing fair AI because it works}}

Our first argument is that computational researchers focus on the CS paradigm of fairness largely ``because it works'', as to quote \citet{knorr-cetina_manufacture_1981}. Conversely, adopting the SOI paradigm necessitates acquiring new material resources and staff, reconfiguring publication venues, and dedicating additional time to novel methodological processes.

In the case of biomedical cancer research, as elaborated in Section \ref{sec:case}, \citet{fujimura_constructing_1987} identifies two facilitators of making a problem doable (and thus, decreasing articulation work). The first facilitator is the leeway provided by \textit{available resources}, such as equipment, space, technology, time, staff, and skills that need to be available for researchers to enable a working routine. Second, a \textit{clear division of labor}, such as clear task-person and task-organization divisions, decreases articulation work, e.g. work of deciding who should do what, when, and how. 

In our study, we identified similar resources as crucial for making the CS paradigm of fairness research work. Much like in Fujimura's observations, computational researchers pragmatically leveraged resources for fairness research just as they would for other research problems. This included acquiring funding and utilizing technologies, staff, and skills in ways similar to how they approached ``other ML problems'' (R5). What partly set fairness research apart from other ML problems, though, was its particular reliance on societal data that presented compelling societal use cases (or `real-world scenarios’). Computational researchers encountered challenges in legitimizing fairness research without access to such data, highlighting that technical resources shaped what was considered a meaningful advancement in the fair AI field by acting as facilitators.

\paragraph{\textbf{Uncertainties and constraints of fair AI}} As \citet{fujimura_constructing_1987} highlights, \textit{uncertainty} increases articulation work as it constrains researchers’ possibilities to plan. In our case study, uncertainties posed constraints on computational researchers. The `projectification' of research, with limited time frames, created general challenges for researchers in planning for the long-term. Information overload (e.g. ‘noise’) added to these uncertainties, as researchers were challenged to decide on their own accord which contributions to the field were important enough to extend, incorporate, or reference in their work. In that regard, the lack of regulatory and technical standards (e.g. benchmarks, toolkits, metrics) for fairness also increased uncertainty. As a result, it required more articulation work to find individualized solutions with the collaborative partners of computational researchers. 

The emergence of new societal actors in the fair AI world added to this uncertainty. Currently, there still exists a strong alignment with more `traditional' actors, such as data-providing private entities and the CS community.
We witnessed that during the alignment with private companies, important issues such as privacy or fairness had been translated and embedded into the specific context of their use, changing or losing their meaning. Eventually, this led to the construction of research problems that might be detached from broader issues that the fair AI world aimed to address, or from what CS researchers themselves considered meaningful. 

\paragraph{\textbf{Making SOI (more) doable}} If uncertainties and burdensome articulation work prevented computational researchers from fully engaging with the SOI paradigm, how can SOI become a `doable problem' for them? From our theoretical standpoint, we now examine this question for each of the three organizational levels.

The doability of a research problem greatly relies on the positions of the \textit{social worlds}. \citet{fujimura_constructing_1987} and \citet{clarke_social_2008} have noted that at times, there may be multiple intersecting social worlds, each with its own interests. This can increase the complexity of the articulation work required. As previously discussed, the fair AI world faces new emerging actors, such as regulatory bodies, becoming more important. Furthermore, different interpretations within social worlds exist regarding what is considered doable. This does not necessarily imply that fairness loses credibility because it is ``empty'' \cite{laclau_85_2001, mackillop_how_2018, kalluri_dont_2020}, but that it can rather hold varying significance for different individuals (e.g., a so-called `boundary object' \cite{star_institutional_1989}). From this perspective, the much-requested call for standards, as sometimes indicated by our respondents, could potentially hinder interdisciplinary collaborations and alignment with diverse social worlds. Perhaps instead of such ethical or regulatory standards on fairness for computational researchers to translate into local (lab) practices, we need more alignment between different social worlds (e.g. CS community, regulatory bodies, companies, marginalized communities). This could include explicit stances from the social worlds’ actors acknowledging the SOI paradigm of fairness and making the SOI perspective an integral part of what is deemed as ``success'' \cite{knorr-cetina_manufacture_1981}.

On the level of the \textit{laboratory}, the criteria of how research is evaluated, what counts as a successful publication, and which venues are treated as important, could be re-evaluated. The success of research is still mostly defined via publications (and within these, the `weight' relies on rigid and decontextualized hierarchies of venue quality). The quest for other forms of visibility is, and will remain, an open problem for both scientometrics theory and academic institutions’ practices. Academic institutions’ ongoing developments of evaluation and indicator systems (which are, to some extent, inherently self-defeating) continues\footnote{e.g., {\href{https://research.kuleuven.be/en/policy-figures/responsible-use-of-metrics}{KU Leuven's position on the responsible use of metrics in the assessment of research(ers).}}}. One potential approach to follow is the purposeful inclusion of SOI-related goals into project proposals, so that project scope and timeline explicitly account for research initiatives such as thorough user studies, to give an example. Following that, it is also important to account for the fact that an enhanced SOI-focus for a project will likely also require a more prominent role for actors from the SOI paradigm within the project structure.

We have made the point that much of what drives the current situation is a need to limit articulation work for overburdened (computational) researchers. Therefore, we acknowledge that suggesting changes on the individual or \textit{experimental} level could actually increase articulation work. However, we need alternative strategies that individual researchers can take if they wish to make impactful changes to the kind of research that gets done. One strategy that we propose is taking a conscious awareness of how we allocate time to various tasks and a thoughtful consideration of how tasks are driven by optimizing productivity and efficiency. From our own experience in collaborating for this paper, we would like to emphasize that interdisciplinary alignment takes a considerable amount of time and that there are misses along the way. To navigate such phases successfully, requires, as one of our interview partners also pointed out, to allow our research to be guided by excitement and curiosity even if that initially seems undoable.

Paradoxically, making SOI research doable could also accentuate, or create, new hierarchies between paradigms and disciplines. Resources are inherently unequal across disciplines and academic institutions \cite{merton_social_2000, larregue_long_2018, albert_boundary-work_2009}. As supported by our findings, this can lead to a scenario where SOI methods and theories are gradually absorbed by computational disciplines that are already situated in a more powerful and financially advantaged position, particularly when it comes to ML research projects. In configuring interdisciplinary projects, close attention to fair resource and influence distributions amongst disciplines and institutions is required. Additionally, there is a need to address open questions about how to make something doable without perpetuating existing hierarchies.

%% file: sections/07limitations.tex
\section{Limitations and Future Research}
\label{sec:limitations}
We hope that our study opens some novel avenues for investigating fair AI from the perspective of the CS researchers involved in it. Yet, our study is limited to only a small group of researchers. While we were especially interested in the perspectives and experiences of these computational researchers, this reduces the scope of our work and hence limits the generalizability of the claims made regarding CS laboratories at large. Future research could integrate the perspectives of other relevant actors in the laboratory. Beyond CS, the doability of fair AI depends on the collaborative efforts of engineers, data scientists, ethicists, research group leaders, or IT experts who may be all situated in the (same) laboratory. Further research could investigate these micro-level actors' interactions with each other, as well as with meso-level actors more closely.

Second, we would find it interesting to further systematically explore our interviews regarding our respondents' definitions of ``fair AI’’ compared to a rather more flexible and open definition adopted while talking about their ongoing research projects and professional routines \cite{jarke_making_2024}. Related to that, we were explicitly interested in how to align CS research with the SOI paradigm, rather than exploring alignment from the reverse perspective. To gain a comprehensive understanding of the current engagements of SOI scholars with fair AI, along with identifying constraints and facilitators, and possible alignment with CS, future research is needed. 

The third limitation pertains to the interview setting employed in our study. While qualitative interviews are valuable for exploring narratives, recounting stories, and uncovering constraints and challenges –- aligning with our research interest –- they also capture retrospective rationalizations of practices and events. Recognizing this, we encourage future research endeavors to conduct (more) ethnographic studies within the laboratory. Engaging in ethnographic studies allows for investigating the doability of tasks in `real-time', and observing activities as they unfold. We think that this will allow for a deeper understanding of the ambiguities of doability, and an exploration of the more tacit aspects of fair ML, which for those we interviewed might have already been ordinary and thus less noteworthy.

Fourth, the advantage of examining researchers' accounts in qualitative settings lies in providing interview participants with the opportunity to articulate ambivalence, correct their perspectives, and even pose questions in return. However, to expand generalizability, future research could broaden the scope of inquiry by incorporating quantitative surveys that could facilitate the collection of data from a larger, and ideally, a more diverse group of participants. Notwithstanding, future research could also consider introducing alternative ways to collect qualitative data, such as the elaboration of hypothetical scenarios (e.g., vignettes, speculative writing) in tandem with our interview partners to reflect upon different real-world situations.

Fifth, we delineated two paradigms, but it is important to consider the possibility of additional paradigms existing, emerging, and changing \cite{clarke_social_2008}. This insight also extends to the actors within the social world and their alignments with the CS lab. Notably, investigating actors excluded from the fair AI world, e.g. identifying who is left out and why, promises deeper insights into unequally distributed power among the social world actors.

As outlined in Section \ref{sec:background}, the intermediate stage of fair AI is linked to an emerging division between SOI and CS among conferences and publication venues. Further exploration of how these divisions specifically affect the research conducted in the CS lab, as already outlined in our study, may hold novel insights about the current and future evolvement of the fair AI world. Conversely, future research might explore how researchers situated in CS labs set out to impact conferences and publication venues to increase alignment with their research, and whether this also includes a shift in their publication choices and conference associations.

It might also be interesting to compare some of our results on the organization of CS research with other areas of academia. For several decades, the project has become the modus operandi of academia \cite{rodder_re-ordering_2012}, and beyond \cite{brockling_entrepreneurial_2016}. According to \citet{boltanski_new_2005}, projects even make up a systematic regime of contemporary (capitalist) societies. Much more research is needed on the neoliberal reorganization of scientific institutions \cite{demirovic_demokratie_2007}, and how the fragmentation of research output into the most possible number of papers becomes part of researchers' work.

Finally, our research findings offer a critical perspective on Fujimura's theory of articulation work, suggesting a need to revisit some of its premises, as our findings shine a light on an important gap: 
the theory focuses on alignment with the social world, but much less on alignment between social world actors with differing and ambiguous stances. Complicating this situation, and as our findings underscore, there is an asymmetrical distribution of power among social world actors. We argue for a deeper exploration of power dynamics within articulation work to elucidate how some exert greater influence in shaping (CS) research agendas and production work than others.

%% file: sections/08conclusion.tex
\section{Outlook}
\label{sec:conclusion}
To understand fair AI research, it is essential to shift our perspective toward its actors, their everyday practices, and their challenges. By investigating the conditions and constraints that CS researchers experience, as well as their frustrations, needs, and desires, we gain important insights into how fair AI is currently made doable, and can derive novel inspirations about \emph{how else it could be}. This points to a need for the consideration of how to embrace more social-oriented and interdisciplinary research in our work and not be confined by what is deemed doable. It also requires exploring avenues that challenge the organizational (and epistemic) status quo.

As CSCW and STS theories remind us, scientific research is, fundamentally, a form of work. Some of the issues we have identified, along with the proposals to address them, mirror long-standing struggles and constructive developments within academic work more broadly. However, engaging in fair AI could also reach beyond the formal boundaries of work. This suggestion is not meant to replace integrating SOI into computational practices but rather serves as a reminder of the diverse opportunities for meaningful engagement with fairness and social justice across various aspects of our lives \cite{dignazio_data_2020,benjamin2022ViralJusticeHow}, surpassing professional responsibilities.